\documentclass[preprint,12pt]{elsarticle}
\usepackage{amssymb}
\usepackage{graphicx}
\usepackage{url}
\def\*#1{\mathbf{#1}}
\usepackage{amsmath}
\journal{Medical Image Analysis}

\begin{document}

\begin{frontmatter}



\title{CUTE-MRI: Conformalized Uncertainty-based framework for Time-adaptivE MRI}


\author[lu,tue]{Paul Fischer} %% Author name
\ead{paul.fischer@uni-tuebingen.de}
\author[tue]{Jan Nikolas Morshuis}
\author[midas,tue]{Thomas Küstner}
\author[lu,tue]{Christian Baumgartner}

\affiliation[lu]{organization={University of Lucerne, Faculty of Health Sciences and Medicine},%Department and Organization
            % addressline={Maria-von-Linden Strasse 6}, 
            city={Lucerne},
            % postcode={72076}, 
            country={Switzerland}}
            
\affiliation[tue]{organization={University of Tübingen,  Cluster of Excellence – Machine Learning for Science},%Department and Organization
            % addressline={Maria-von-Linden Strasse 6}, 
            city={Tübingen},
            % postcode={72076}, 
            country={Germany}}
            
\affiliation[midas]{organization={University Hospital of Tübingen,  Medical Image and Data Analysis Lab},%Department and Organization
            % addressline={Maria-von-Linden Strasse 6}, 
            city={Tübingen},
            % postcode={72076}, 
            country={Germany}}

%% Abstract
\begin{abstract}
Magnetic Resonance Imaging (MRI) offers unparalleled soft-tissue contrast but is fundamentally limited by long acquisition times. While deep learning-based accelerated MRI can dramatically shorten scan times, the reconstruction from undersampled data introduces ambiguity resulting from an ill-posed problem with infinitely many possible solutions that propagates to downstream clinical tasks. This uncertainty is usually ignored during the acquisition process as acceleration factors are often fixed \textit{a priori}, resulting in scans that are either unnecessarily long or of insufficient quality for a given clinical endpoint. This work introduces a dynamic, uncertainty-aware acquisition framework that adjusts scan time on a per-subject basis. Our method leverages a probabilistic reconstruction model to estimate image uncertainty, which is then propagated through a full analysis pipeline to a quantitative metric of interest (e.g., patellar cartilage volume or cardiac ejection fraction). We use conformal prediction to transform this uncertainty into a rigorous, calibrated confidence interval for the metric. During acquisition, the system iteratively samples k-space, updates the reconstruction, and evaluates the confidence interval. The scan terminates automatically once the uncertainty meets a user-predefined precision target. We validate our framework on both knee and cardiac MRI datasets. Our results demonstrate that this adaptive approach reduces scan times compared to fixed protocols while providing formal statistical guarantees on the precision of the final image. This framework moves beyond fixed acceleration factors, enabling patient-specific acquisitions that balance scan efficiency with diagnostic confidence, a critical step towards personalized and resource-efficient MRI.
\end{abstract}


% \begin{highlights}
% \end{highlights}

%% Keywords

\end{frontmatter}

% \linenumbers


\section{Introduction}
\label{Intro}
Magnetic Resonance Imaging (MRI) is a cornerstone of modern medical diagnostics. Its ability to non-invasively generate images with exceptional soft-tissue contrast makes it indispensable for the diagnosis, staging, and monitoring of a wide range of diseases, from neurological disorders to musculoskeletal injuries and cardiovascular conditions \cite{westbrook2018mri}. However, the high diagnostic value of MRI is often counterbalanced by its inherently long acquisition times. These lengthy scans can lead to patient discomfort, increase the likelihood of motion artifacts that degrade image quality, and limit patient throughput, thereby increasing operational costs and wait times \cite{andre2015toward}. Consequently, accelerated MRI techniques, which aim to reconstruct high-quality images from undersampled k-space data, are of paramount importance for making MRI more efficient, cost-effective, and patient-friendly \cite{lustig_sparse_2007,knoll2020deep}.

While accelerated MRI promises to alleviate these challenges, the majority of current methods, both in clinical practice and in research, rely on static acquisition strategies \cite{lustig_sparse_2007,knoll2020deep,jaspan2015compressed}. These approaches employ fixed, pre-determined undersampling rates that are designed offline and are not adapted to the specific patient. This inflexibility represents a central, unaddressed limitation: the acquisition process remains agnostic to the content and complexity of the image being formed. This can lead to a suboptimal use of scanner time, as less data may be sufficient, especially when a specific downstream metric is the primary interest. 

The evolution of accelerated MRI has been marked by two major paradigms. The first encompasses classic reconstruction techniques, such as parallel imaging and compressed sensing, while the second is defined by the rise of deep learning (DL). Classic methods, rooted in parallel imaging (e.g., SENSE, GRAPPA) and compressed sensing (CS), leverage explicit priors like signal sparsity to recover images from limited data \cite{lustig_sparse_2007,pruessmann1999sense,griswold2002generalized}. While they provide a strong theoretical foundation, their performance tends to degrade at high acceleration factors, where severe aliasing artifacts can become diagnostically prohibitive. In contrast, the second paradigm of deep learning has revolutionized the field. Models trained on large datasets learn complex, implicit priors and have demonstrated high-quality reconstructions, even from highly undersampled data \cite{hammernik2018learning,schlemper2017deep,hammernik2023physics,heckel2024deep}. These methods often outperform traditional techniques in terms of pure reconstruction quality and speed.

Despite their impressive performance, deep learning models often function as "black boxes," and their predictions come with no inherent guarantees of correctness. This can lead to a critical problem of misplaced trust, where models may produce plausible-looking but factually incorrect reconstructions, a phenomenon often termed "hallucination" \cite{maynez2020faithfulness,aithal2024understandinghallucinationsdiffusionmodels,antun2020on}. The risk is particularly acute in the ill-posed problem of MR reconstruction, where uncertainty arises not only from the missing k-space measurements but also from physiological and anatomical variability between patients, pathologies, and motion.  This unquantified uncertainty does not just affect the reconstructed image; it can silently propagate to and corrupt downstream clinical tasks, such as segmentation, registration, or disease classification, that rely on these images for diagnosis and treatment planning \cite{fischer2023uncertainty,wundram2024leveraging}.

Recognizing this challenge, a growing body of research has focused on uncertainty quantification (UQ) for deep learning in medical imaging. Various methods, such as Bayesian neural networks, ensembles and variational autoencoder-based methods have been developed to estimate model uncertainty \cite{gal2016dropout,lakshminarayanan2017simple,phiseg,edupuganti2020uncertainty,narnhofer2021bayesian,huttinga2023gaussian,schlemper2018bayesian,morshuis2024segmentation,morshuis2025minddetail}. Several works have successfully demonstrated how this uncertainty can be propagated from the reconstruction to a downstream task to provide a more complete picture of diagnostic confidence \cite{fischer2023uncertainty,wundram2024conformal}.

However, while significant research has focused on estimating and propagating uncertainty for post-hoc analysis, its potential to actively guide and optimize the MRI acquisition process itself in real-time remains largely unexplored. Daudé et al. \cite{daude2024inline} proposed an adaptive method where scan quality, specifically the Signal-to-Noise Ratio SNR, is estimated periodically during acquisition. The scan is terminated once the SNR surpasses a pre-defined quality threshold, enabling personalized scan durations. However, this approach relies on a classical, signal-based metric and does not account for the reconstruction uncertainty or potential for artifacts, such as hallucinations, common in modern learning-based methods. Pineda et al. \cite{pineda2020active} for example analyzed how to find the optimal sampling trajectory for accelerated MR acquisition using reinforcement learning, however they did not consider the effect of downstream applications. Wang et al. \cite{wang2024promoting} jointly analyzed the influence of k-space acquisition and segmentation quality by iteratively sampling k-space up to a fixed undersampling rate such that segmentation quality is as high as possible. However, this work does not account for the inherent uncertainty in the pipeline, nor does it assess when there is sufficient k-space data.. It becomes apparent that prior work on optimizing scan duration has typically focused on pre-calculating sampling trajectories or defining stopping criteria based on image-level metrics, without considering model confidence along the diagnostic pipeline \cite{huang2025udnet,wu2024learning}. This reveals a critical gap: current static acquisition protocols are inherently inefficient. They may waste valuable scanner time on anatomically ``easy'' cases that could have been reconstructed with sufficient quality from fewer measurements, or conversely, they may terminate prematurely for ``hard'' or unusual cases, yielding diagnostically inadequate images. This one-size-fits-all approach fails to account for the simple fact that some diagnostic tasks or anatomies do not require perfectly reconstructed images to yield clinically reliable results.

In this work, we show that by monitoring the uncertainty of a reconstruction model and its downstream clinical application, one can create a patient-specific, adaptive stopping rule for k-space acquisition. The core idea is to halt the scan precisely when the system reaches a pre-defined level of diagnostic confidence, rather than adhering to a fixed sampling budget. Such a dynamic stopping criterion would optimize the scan duration for each individual, allowing for fast scan times while keeping the diagnostic quality high. This would not only improve patient comfort and scanner throughput but would do so without sacrificing the diagnostic integrity required for clinical decision-making.

To this end, we introduce CUTE-MRI: a Conformalized Uncertainty-based framework for Time-adaptivE MRI. This novel framework leverages uncertainty estimation to determine an optimal, patient-specific stopping point for the scan, ensuring that the resulting images are fit for a specified clinical purpose. Our main contributions are threefold:

\begin{enumerate}
    \item We propose a complete framework for dynamically terminating an MR acquisition based on the propagation of uncertainty through a diagnostic pipeline, from reconstruction to a downstream clinical measurement.
    \item We demonstrate that na\"ive uncertainty estimates from deep learning models without adjustment are poorly calibrated and thus unsuitable for reliable decision-making. To address this, we show how to transform these estimates into rigorous confidence intervals with formal statistical guarantees using the principled technique of conformal prediction.
    \item We validate our framework on two distinct and clinically relevant applications: the estimation of patellar cartilage volume from knee MRI and the computation of left ventricular ejection fraction from cardiac CINE MRI, demonstrating its effectiveness and generalizability.
\end{enumerate}

\section{Methods} \label{sec:methods}
\begin{figure}[t]
    \centering
    \includegraphics[width=0.99\linewidth]{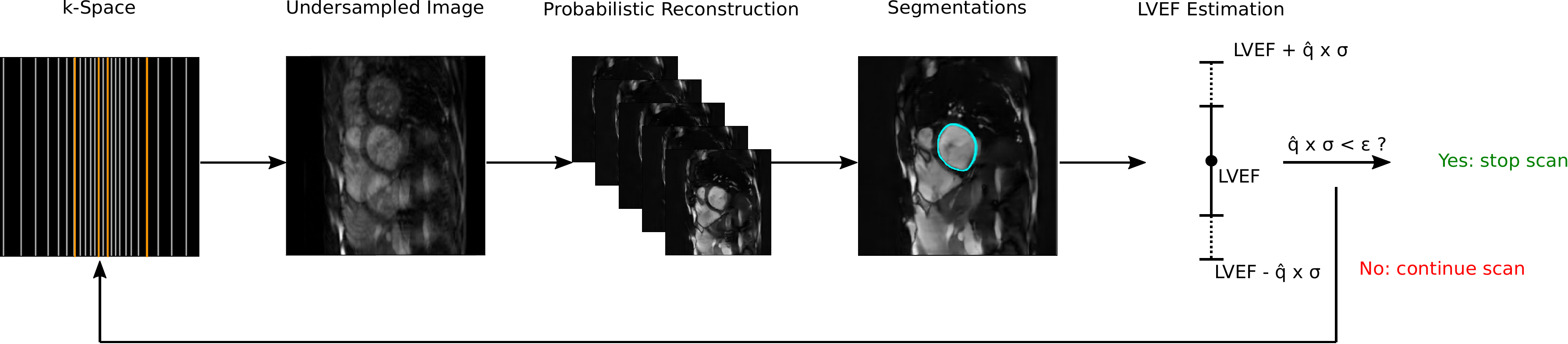}
    \caption{Overview of the proposed dynamic and iterative MR acquisition framework. At each time step $t$, k-space data $\mathbf{y}_t$ is acquired. A probabilistic model generates $M$ candidate reconstructions $\{\mathbf{x}^{(m)}\}$, which are then passed to a segmentation network. The resulting segmentations are used to compute a distribution of a clinical metric (e.g., LVEF). The uncertainty of this metric is estimated and calibrated. Based on a user-defined stopping criterion (i.e., if the uncertainty is below a threshold $\varepsilon$), the scan is either terminated or continued with the acquisition of the next k-space segment.}
    \label{fig:method}
\end{figure}

We propose a dynamic acquisition pipeline that iterates over a set of undersampling rates, assesses the uncertainty of derived clinical metrics and stops the scan once a predefined confidence threshold is reached. The pipeline operates as follows: after each k-space acquisition step, we first generate a set of $M$ plausible reconstructions $\{\mathbf{x}^{(m)}\}_{m=1}^M$ from the currently undersampled k-space data $\mathbf{y}_t$ using a probabilistic reconstruction model, PHiRec \cite{fischer2023uncertainty}, which we describe in Section \ref{sec:phirec}. In Section \ref{sec:propag} we showcase how to propagate uncertainty where each candidate reconstruction $\mathbf{x}^{(m)}$ is segmented by a deterministic segmentation network, $S(\cdot)$, yielding a set of segmentations $\{ \mathbf{s}^{(m)} \}_{m=1}^M$, where $\mathbf{s}^{(m)} = S(\mathbf{x}^{(m)})$. From these segmentations, a clinical metric of interest, $\mathbf{w}$, is computed via a function $f(\cdot)$, resulting in a set of metric samples $\{ \mathbf{w}^{(m)} \}_{m=1}^M$, where $\mathbf{w}^{(m)} = f(\mathbf{s}^{(m)})$. In our experiments, these metrics are the left ventricular ejection fraction and patellar cartilage volume. We quantify the uncertainty of the metric $\mathbf{w}$ by its empirical standard deviation, which is then calibrated using a scaling factor derived from conformal prediction (Section \ref{sec:conf}). This entire process—reconstruction, segmentation, metric estimation, and uncertainty calibration—is repeated after each acquisition step. The acquisition is terminated when the calibrated uncertainty bound falls below a user-defined threshold, $\varepsilon$. A schematic of this iterative process is provided in Figure \ref{fig:method}.

\subsection{Probabilistic Hierarchical Reconstruction (PHiRec)}
\label{sec:phirec}

The goal of MR reconstruction is to recover a high-fidelity image $\mathbf{x} \in \mathbb{C}^D$ from undersampled k-space measurements $\mathbf{y} \in \mathbb{C}^M$, where $M \ll D$. The relationship is described by the forward model:
\begin{equation}
    \mathbf{y} = \mathcal{A}(\mathbf{x}) + \mathbf{n} = \mathcal{M}\mathcal{F}\mathcal{S}\mathbf{x} + \mathbf{n},
    \label{eq:forward_model}
\end{equation}
where $\mathcal{S}$ denotes the coil sensitivity mapping, $\mathcal{F}$ is the Fourier transform, $\mathcal{M}$ is the binary sampling mask, and $\mathbf{n}$ represents measurement noise. The combined operator $\mathcal{A}$ is the forward encoding model.

Instead of seeking a single point estimate, we aim to model the full posterior distribution $p(\mathbf{x} \mid \mathbf{y})$. This inverse problem can be framed as a de-aliasing task by conditioning on the zero-filled reconstruction $\mathbf{x}_u = \mathcal{A}^*(\mathbf{y})$, where $\mathcal{A}^*$ is the adjoint of the forward operator. We thus seek to model the distribution $p(\mathbf{x} \mid \mathbf{x}_u)$.

To this end, we employ our previously proposed Probabilistic Hierarchical Reconstruction (PHiRec) model \cite{fischer2023uncertainty}, a state-of-the-art method for uncertainty quantification in MR reconstruction. Its high sampling speed, compared to alternatives like diffusion models, makes it particularly suitable for the real-time requirements of our dynamic acquisition setting. PHiRec is a hierarchical conditional variational autoencoder (CVAE) that models the distribution of reconstruction artifacts across multiple scales. It uses a hierarchy of latent variables $\mathbf{z}_{1:L} = \{\mathbf{z}_1, \dots, \mathbf{z}_L\}$, where each level $l$ corresponds to a different image resolution. The generative process is defined as:
\begin{equation}
    p(\mathbf{x} | \mathbf{x}_u) = \int p(\mathbf{x} | \mathbf{z}_1, \mathbf{x}_u) \left( \prod_{l=1}^{L-1} p(\mathbf{z}_l | \mathbf{z}_{l+1}, \mathbf{x}_u) \right) p(\mathbf{z}_L | \mathbf{x}_u) \, d\mathbf{z}_{1:L}.
    \label{eq:phirec_distr}
\end{equation}
The model is trained by maximizing the evidence lower bound (ELBO) on the log-likelihood of the data, which, for a given ground truth image $\mathbf{x}$, is formulated as:
\begin{align}
    \mathcal{L}_{\text{ELBO}}(\mathbf{x}, \mathbf{x}_u) = & \mathbb{E}_{q(\mathbf{z}_{1:L}|\mathbf{x}, \mathbf{x}_u)}[\log p(\mathbf{x}|\mathbf{z}_{1:L}, \mathbf{x}_u)] \nonumber \\ 
    & - \sum_{l=1}^L \text{KL}\left(q(\mathbf{z}_l|\mathbf{z}_{>l}, \mathbf{x}, \mathbf{x}_u) \parallel p(\mathbf{z}_l|\mathbf{z}_{>l}, \mathbf{x}_u)\right).
    \label{eq:elbo}
\end{align}
Here, $q(\cdot)$ is the approximate posterior (encoder) and $p(\cdot)$ is the prior (decoder). Assuming Gaussian distributions for the likelihood and the latent priors, maximizing the ELBO is equivalent to minimizing a loss function composed of two main terms: a reconstruction loss (typically mean squared error) corresponding to the first term, and a regularization term that penalizes the divergence between the approximate posterior and the prior distributions for each latent level, given by the sum of KL-divergences.

\subsubsection{Segmentation}
For the downstream segmentation task, we employed a standard 2D U-Net architecture \cite{ronneberger2015u}. The network follows a symmetric encoder-decoder structure with four downsampling stages. The encoder path begins with an initial block of two 3×3 convolutions, mapping the input channels to 64 feature maps. Each subsequent downsampling stage consists of a 2×2 max-pooling operation followed by two more 3×3 convolutions, doubling the number of feature channels at each step (64 → 128 → 256 → 512 → 1024).

The decoder path symmetrically mirrors this design. At each stage, it uses a 2×2 transposed convolution to upsample the feature maps, followed by concatenation with the corresponding feature maps from the encoder path via skip connections. These concatenated features are then processed by two 3×3 convolutions. All convolutional layers, except for the final one, are followed by Batch Normalization and a ReLU activation function. A final 1×1 convolution maps the 64 feature channels from the last upsampling block to the number of output classes, producing the segmentation logits. The model was trained with the fully sampled reconstructions as input, using a hybrid loss function, defined as the sum of a soft Dice loss ($\mathcal{L}_{\text{Dice}}$) and a standard Cross-Entropy loss ($\mathcal{L}_{\text{CE}}$):
\begin{equation}
    \mathcal{L}_{\text{seg}} = \mathcal{L}_{\text{Dice}} + \mathcal{L}_{\text{CE}}
    \label{eq:seg_loss}
\end{equation}

\subsection{Uncertainty Propagation through the Processing Pipeline}
\label{sec:propag}
To quantify how uncertainty from the reconstruction stage affects downstream clinical metrics, we propagate samples through the entire analysis pipeline. This Monte Carlo approach allows us to estimate the posterior distribution of a given metric, conditioned on the undersampled k-space data.

Let $\mathbf{y}$ denote the undersampled k-space measurements for a given scan. Our probabilistic reconstruction network is trained to sample from the posterior distribution of the fully-sampled image, $p(\mathbf{x}|\mathbf{y})$. For each $\mathbf{y}$, we draw a set of $M$ plausible image reconstructions:
\begin{equation}
    \{ \hat{\mathbf{x}}^{(m)} \}_{m=1}^M \sim p(\mathbf{x}|\mathbf{y}) \text{,}
\end{equation}
where each $\hat{\mathbf{x}}^{(m)}$ is a sample. Let $T(\cdot)$ be a deterministic function representing a downstream task (e.g., segmentation followed by volume calculation) that computes a scalar metric of interest, $\mathbf{w}$. By applying this function to each reconstruction sample, we generate a set of metric samples:
\begin{equation}
    \{ \mathbf{w}^{(m)} = T(\hat{\mathbf{x}}^{(m)}) \}_{m=1}^M \text{.}
\end{equation}
These samples, $\{\mathbf{w}^{(m)}\}$, form an empirical estimate of the metric's posterior distribution, $p(\mathbf{w}|\mathbf{y})$. From this set, we can compute the final prediction as the sample mean, $\hat{\mathbf{w}}$, and an estimate of its uncertainty as the sample standard deviation, $\sigma_{\mathbf{w}}$:
\begin{equation}
    \hat{\mathbf{w}} = \frac{1}{M} \sum_{m=1}^M \mathbf{w}^{(m)}, \quad \quad \sigma_{\mathbf{w}} = \sqrt{\frac{1}{M-1} \sum_{m=1}^M (\mathbf{w}^{(m)} - \hat{\mathbf{w}})^2}
\end{equation}
This allows us to define a one-standard-deviation interval, $\mathcal{I}_{\text{std}} = [\hat{\mathbf{w}} - \sigma_{\mathbf{w}}, \hat{\mathbf{w}} + \sigma_{\mathbf{w}}]$. A smaller interval suggests a more certain prediction.

For both datasets, the function $T(\cdot)$ involves applying a trained segmentation network, $S(\cdot)$, to the reconstruction samples. For each subject, we generate $M=20$ reconstructions, yielding a set of $M$ segmentation masks $\{\hat{\mathbf{s}}^{(m)} = S(\hat{\mathbf{x}}^{(m)})\}_{m=1}^M$. These masks are then used to compute the final clinical metrics.

\subsection{Uncertainty Calibration via Conformal Prediction}
\label{sec:conf}
While the standard deviation $\sigma_{\mathbf{w}}$ provides a useful heuristic for uncertainty, the resulting intervals lack formal statistical guarantees. To construct prediction intervals with rigorous theoretical properties, we employ the split conformal prediction framework \cite{vovk2005algorithmic, angelopoulos2021gentle}. This method transforms heuristic uncertainty estimates into valid prediction intervals that are guaranteed to contain the true, unknown value with a user-specified probability.

Formally, for a new test sample with undersampled data $y$, we aim to construct a prediction interval $\mathcal{C}(\mathbf{y})$ for the true metric $\mathbf{w}$ that satisfies the marginal coverage guarantee:
\begin{equation}
    \mathbb{P}(\mathbf{w} \in \mathcal{C}(\mathbf{y})) \ge 1 - \alpha
    \label{eq:cp_guarantee}
\end{equation}
where $\alpha \in (0, 1)$ is a user-defined tolerable error rate. This procedure requires a dedicated calibration set $D_{\text{calib}} = \{(\mathbf{y}_i, \mathbf{w}_i)\}_{i=1}^{n_{\text{calib}}}$, where samples are assumed to be exchangeable with the test data.

The core idea is to define a nonconformity score that quantifies how "unusual" a prediction is, given our heuristic uncertainty. For our symmetric intervals based on the standard deviation, we define the score for each calibration sample $i$ as the normalized absolute error:
\begin{equation}
    sc_i = \frac{|\mathbf{w}_i - \hat{\mathbf{w}}_i|}{\sigma_{\mathbf{w},i}}
    \label{eq:score_function}
\end{equation}
where $\hat{\mathbf{w}}_i$ and $\sigma_{\mathbf{w},i}$ are the mean prediction and standard deviation derived from the Monte Carlo samples for calibration sample $i$ and $\mathbf{w}_i$ is the ground truth value. These scores $\{sc_i\}_{i=1}^{n_{\text{calib}}}$ measure the error in units of predicted standard deviations.

We then compute a correction factor, $\hat{q}$, by taking the $\lceil(1-\alpha)(n_{\text{calib}}+1)\rceil$-th value of the sorted nonconformity scores. This $\hat{q}$ represents the empirical quantile of the normalized errors on the calibration set. The final conformal prediction interval for each new test prediction $(\hat{\mathbf{w}}, \sigma_{\mathbf{w}})$ is then constructed by scaling the standard deviation by this factor:
\begin{equation}
    \mathcal{C}(\mathbf{y}) = [ \hat{\mathbf{w}} - \hat{q}\sigma_{\mathbf{w}}, \hat{\mathbf{w}} + \hat{q}\sigma_{\mathbf{w}} ]
    \label{eq:conformal_interval}
\end{equation}
By construction, this interval is guaranteed to achieve the coverage defined in Eq. \eqref{eq:cp_guarantee}. The width of this interval provides a rigorous, data-driven measure of uncertainty. A wider interval indicates that a larger deviation from the prediction is needed to be considered "conformal," implying higher uncertainty and a greater probability of a large error. This property makes these intervals highly suitable for defining an uncertainty-based stopping criterion for accelerated MRI.

\section{Experiments}

We demonstrate the dynamic uncertainty-guided MR acquisition strategy described in Section~\ref{sec:methods} on two datasets that provide raw multi-coil k-space data: The public Stanford Knee MRI Multi-Task Evaluation (SKM-TEA) \cite{desai2022skm}, and an in-house cardiac CINE MR dataset. We simulate the acquisition process by retrospectively undersampling the k-space data. The two datasets contain anatomical segmentations which allow training a segmentation network and quantify anatomical volumes as introduced earlier, as well as evaluating the method. In the following we describe the experimental setup and the experimental details. 

\subsection{Experimental Setup}
To simulate dynamic MR acquisition, we retrospectively undersample the fully-sampled raw k-space data using predefined sampling masks corresponding to various acceleration factors $R \in {4, 8, \dots, 32}$, as described in section \ref{sec:data}. Starting from a highly undersampled input, we incrementally reveal additional k-space data by successively applying sampling masks of increasing density. The acquisition simulation proceeds by moving to the next predefined k-space subset at each acquisition step, mimicking a real-time, progressive acquisition process. At each step, we generate a reconstruction sample and compute the downstream metric of interest (i.e., patellar cartilage volume or LVEF) along with a calibrated uncertainty interval as described above. The scan is automatically terminated when the uncertainty interval for the downstream metric becomes sufficiently tight—i.e., once the width of the interval falls below a user-defined threshold $\varepsilon$. For the patellar cartilage volume, we defined $\varepsilon_v = 0.5 cm^3$ and for the LVEF as $\varepsilon_{LVEF} = 15 \%$. The code will be available at \url{https://github.com/paulkogni/CUTE-MRI}.

\subsection{Data and Preprocessing}
\label{sec:data}
% We utilized one publicly available dataset, the Stanford Knee MRI Multi-Task Evaluation (SKM-TEA) \cite{desai2022skm}, and an in-house cardiac CINE MRI dataset. Both datasets provide raw multi-coil k-space data and corresponding ground truth segmentations. This multi-task setup enables the evaluation of reconstruction quality and downstream segmentation performance, which is essential for our proposed method of dynamically determining an optimal stopping point.

\subsubsection{SKM-TEA}
The SKM-TEA dataset provides raw multi-coil k-space measurements of knee MRIs, accompanied by manual segmentations of six anatomical structures. While the original dataset includes undersampling masks for up to 16x acceleration based on a Poisson-Disc sampling pattern, we generated a new set of masks to explore higher acceleration factors. We followed the same sampling methodology to create masks for a set of acceleration factors $R \in \{4, 8, 12, 16, 20, 24, 28, 32\}$. The input images for our models were obtained by applying the adjoint operator ($\mathcal{A}^{*}$) to the zero-filled, retrospectively undersampled multi-coil k-space data. As in the original dataset, consistent spatial dimension across subjects was ensured by zero-padding the undersampled k-space.

For our experiments, a dedicated calibration set was required. We created this set by reallocating five subjects from the original training set and five from the original validation set. The test set remained unchanged, as defined by the original benchmark. This partitioning resulted in final splits of 81, 28, 10, and 36 subjects for training, validation, calibration, and testing, respectively.

\subsubsection{CINE}
Our in-house CINE dataset comprises raw multi-coil k-space measurements from cardiac MRI scans, with corresponding manual segmentations for the left ventricle (LV), myocardium (Myo), and right ventricle (RV). Multi-slice 2D Cartesian data was acquired with a balanced steady-state free precession (bSSFP) CINE (2x GRAPPA accelerated) in 8 breath-holds of 12s duration (2 slices per breathhold) each with 20 seconds pause in between. Further imaging parameters include 1.9×1.9mm in-plane (acquired and reconstructed) resolution, slice thickness 8 mm, temporal resolution ~40 ms, 25 cardiac phases (reconstructed), TE=1.06ms, TR=2.12ms, flip angle 52°, bandwidth=915Hz/px. 

Due to the dynamic nature of the CINE acquisition, we employed a Variable-density Incoherent Spatio-Temporal Acquisition (VISTA) sampling pattern \cite{ahmad2015variable} to generate the retrospective undersampling masks. Masks were generated for the same set of acceleration factors $R$ as used for the SKM-TEA dataset. Similarly, input images were reconstructed by applying the adjoint operator ($\mathcal{A}^{*}$)) to the zero-filled multi-coil k-space data. Like for the SKM-TEA dataset, consistent spatial dimension across subjects was ensured by zero-padding the undersampled k-space.

The full CINE cohort includes 134 subjects suitable for the reconstruction task. A subset of 40 subjects has corresponding ground truth segmentations (manually annotated by experienced radiologists with $>10$ years of experience in cardiovascular imaging), enabling the segmentation task. This disparity required us to define two distinct data splits. To ensure a fair comparison and prevent data leakage, the test and calibration sets were kept consistent across both splits.
\begin{itemize}
    \item \textbf{Reconstruction Task:} The 134 subjects were partitioned into 95 for training, 24 for validation, and 10 for testing.
    \item \textbf{Segmentation Task:} The 40 subjects with annotations were split into 20 for training, 5 for validation, 5 for calibration, and the same 10 for testing.
\end{itemize}

\subsection{Training Procedures}
This section outlines the training protocols for the reconstruction and segmentation models. For reproducibility, we maintained consistent hyperparameters where appropriate and detail any dataset-specific adaptations.

\subsubsection{Reconstruction}
A separate PHiRec model was trained for each dataset and acceleration factor $R \in \{4, 8, \dots, 32\}$. The models operate on 2D complex-valued image slices, which are processed as two-channel real-valued tensors corresponding to the real and imaginary part ($\mathbb{R}^{H \times W \times 2}$), where $H$ and $W$ represent the image height and width. For both datasets, the images were normalized per-slice as in the original PHiRec paper~\cite{fischer2023uncertainty}. 

The model architecture was adapted to the different spatial dimensions of the datasets: $512 \times 512$ for SKM-TEA and $192 \times 192$ for CINE. This was achieved by setting the number of resolution levels in the PHiRec network to seven for SKM-TEA and five for CINE. All other model parameters were kept consistent. 

We trained each reconstruction model using the Adam optimizer \cite{kingma2014adam} with a learning rate of $1 \times 10^{-4}$ and a batch size of 12. To improve generalization, we applied spatial data augmentation in the form of random flips and rotations. Training was performed for a fixed duration of 10 days on a single NVIDIA A100 GPU, which was sufficient to ensure convergence. For each acceleration factor, we selected the model checkpoint that achieved the highest Structural Similarity Index (SSIM) \cite{wang2004image} on the validation set for final evaluation.

\subsubsection{Segmentation}
The segmentation U-Net was trained on fully-sampled, normalized 2D image slices with spatial dimensions of $512 \times 512$ for SKM-TEA and $192 \times 192$ for CINE. We used the Adam optimizer with a learning rate of $1\times10^{-4}$ and a batch size of 12. Also here, we used random flips and rotations to increase generalization and model robustness. Training was performed on NVIDIA RTX 2080Ti GPUs. The models were trained for maximally three days to ensure convergence. The final model for each dataset was selected based on the checkpoint that achieved the highest mean Dice Similarity Coefficient (DSC) on the validation set.

\subsection{Downstream Metrics and Uncertainty Quantification}
\subsubsection{Patellar Cartilage Volume for SKM-TEA}
For the SKM-TEA dataset, we used the patellar cartilage volume as our downstream metric, as it is recognized as a biomarker for osteoarthritis \cite{ciliberti2022role}. We defined a function $V(\cdot)$ that calculates the volume from a segmentation mask in cm$^3$, using the voxel spacing provided in the image metadata. This yields a set of volume samples $\{v^{(m)} = V(\hat{\mathbf{s}}^{(m)})\}_{m=1}^{M}$. From these samples, we compute the final volume prediction, $\hat{v}$, and its associated uncertainty, $\sigma_v$.

\subsubsection{Ejection Fraction for CINE}
For the CINE dataset, the metric of interest was the Left Ventricular Ejection Fraction (LVEF), a critical biomarker for cardiac function. Calculating LVEF requires segmenting the left ventricle at two specific cardiac phases: end-diastole (ED) and end-systole (ES).

For each subject, we generate $20$ reconstruction samples for both the ED scan, $\{\hat{\mathbf{x}}_{\text{ED}}^{(m)}\}$, and the ES scan, $\{\hat{\mathbf{x}}_{\text{ES}}^{(m)}\}$. We then apply the segmentation network to each, obtaining paired sets of segmentation masks: $\{\hat{\mathbf{s}}_{\text{ED}}^{(m)}\}$ and $\{\hat{\mathbf{s}}_{\text{ES}}^{(m)}\}$. The corresponding ED and ES volumes, $v_{\text{ED}}^{(m)}$ and $v_{\text{ES}}^{(m)}$, are calculated for each possible pairing. This yields us $M=20 \times 20 = 400$ LVEF samples using its clinical definition:
\begin{equation}
    \text{LVEF}^{(m)} = \frac{v_{\text{ED}}^{(m)} - v_{\text{ES}}^{(m)}}{v_{\text{ED}}^{(m)}} \times 100\%
    \label{eq:lvef}
\end{equation}
This process yields an empirical distribution of LVEF values, from which we compute the final prediction, $\hat{\text{LVEF}}$, and its uncertainty, $\sigma_{\text{LVEF}}$.

While the standard deviation $\sigma_{\mathbf{w}}$ provides an intuitive measure of uncertainty, the resulting interval $\mathcal{I}_{\text{std}}$ offers no formal guarantees on its coverage probability (i.e., how often it contains the true, unknown metric value). To construct prediction intervals with rigorous statistical guarantees, we leverage the conformal prediction framework, as detailed in the following section.

\subsubsection{Calibration Details}
For all calibration experiments, as described in Section \ref{sec:conf}, we set the target error rate to $\alpha = 0.1$, aiming for 90\% coverage. The calibration procedure was performed independently for each acceleration factor $R$. This was done using the dedicated calibration sets described previously, with $n_{\text{calib}}=10$ for SKM-TEA and $n_{\text{calib}}=5$ for CINE.

\section{Results}
After training the models and calibrating the uncertainties as described above, we evaluated our proposed framework in three steps. First, we quantified the performance of the underlying reconstruction and segmentation models in Section \ref{sec:rec_seg}. In Section \ref{subsec:dyn_stop}, we analyzed the behavior of the dynamic stopping mechanism, comparing outcomes with and without uncertainty calibration. Finally, we present qualitative examples to visualize the method's performance in Section\ref{sec:qualitative}.

\subsection{Reconstruction and Segmentation Performance}
\label{sec:rec_seg}
\begin{figure}[t]
    \centering
    \includegraphics[width=0.99\linewidth]{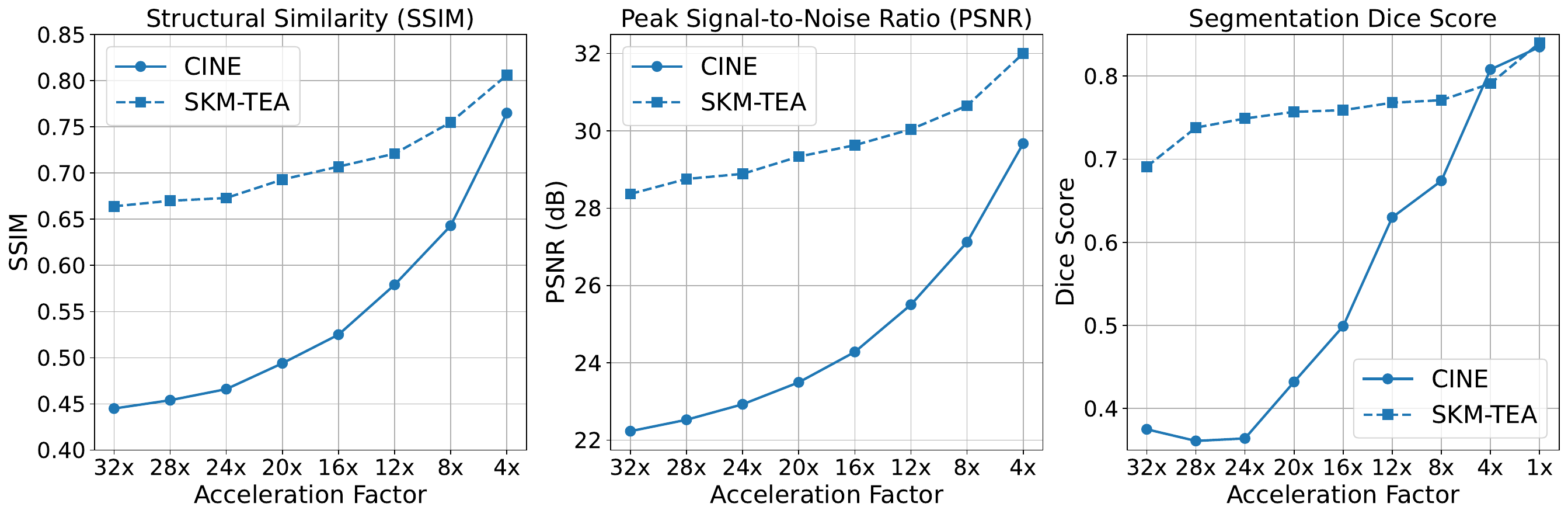}
    \caption{Quantitative evaluation of reconstruction and segmentation performance across different acceleration factors for two datasets (SKM-TEA and CINE). Each subplot shows one metric: Structural Similarity Index Measure (SSIM), Peak Signal-to-Noise Ratio (PSNR), and Segmentation Dice Score. The x-axis denotes the acceleration factor (higher values correspond to stronger undersampling). Performance consistently improves with decreasing acceleration, where the models for SKM-TEA yield better metrics compared to the models for the CINE dataset due to a differences in undersampling.}
    \label{fig:metrics}
\end{figure}

To validate the underlying models, we evaluated reconstruction quality using the Structural Similarity Index (SSIM) and Peak Signal-to-Noise Ratio (PSNR), and segmentation accuracy using the Dice Similarity Coefficient (DSC). For the DSC computation, we first calculated the average DSC score across all segmented structures for each patient, and then averaged these scores across all patients. Figure~\ref{fig:metrics} shows that for both datasets, all metrics improved as the acceleration rate decreased, with the highest scores achieved in the fully-sampled setting. This trend is expected, as more k-space data provides more information for both reconstruction and the downstream segmentation task.

We also observed that performance on the CINE dataset was notably lower than on the SKM-TEA dataset across all acceleration factors. This difference can be attributed to the more challenging VISTA undersampling pattern used for the CINE data, which tends to produce stronger aliasing artifacts in zero-filled images compared to the Poisson-disk sampling used for SKM-TEA.

\subsection{Dynamic Stopping Behavior and Coverage}
\label{subsec:dyn_stop}
\begin{figure}[t]
    \centering
    \includegraphics[width=0.49\linewidth]{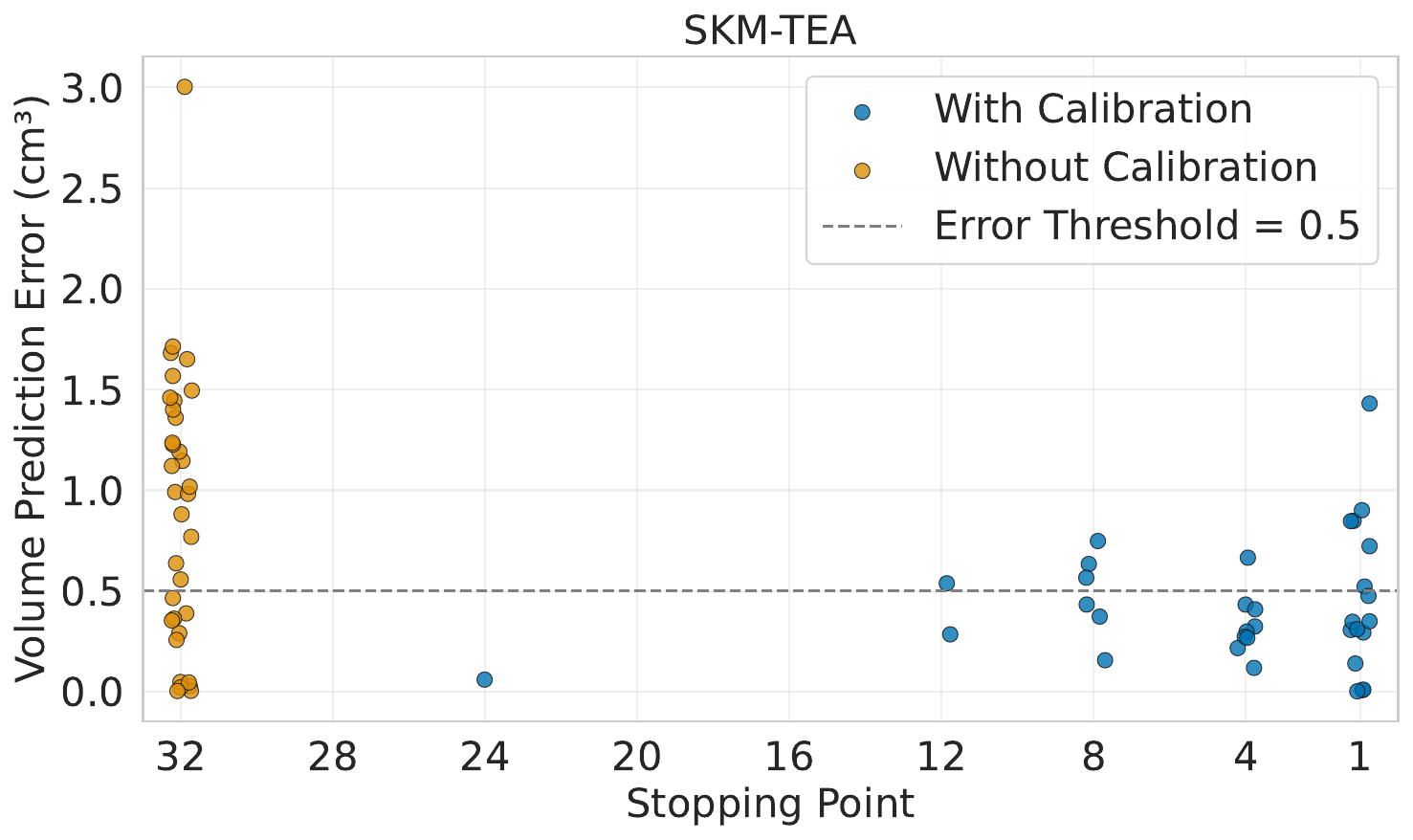}
    \includegraphics[width=0.49\linewidth]{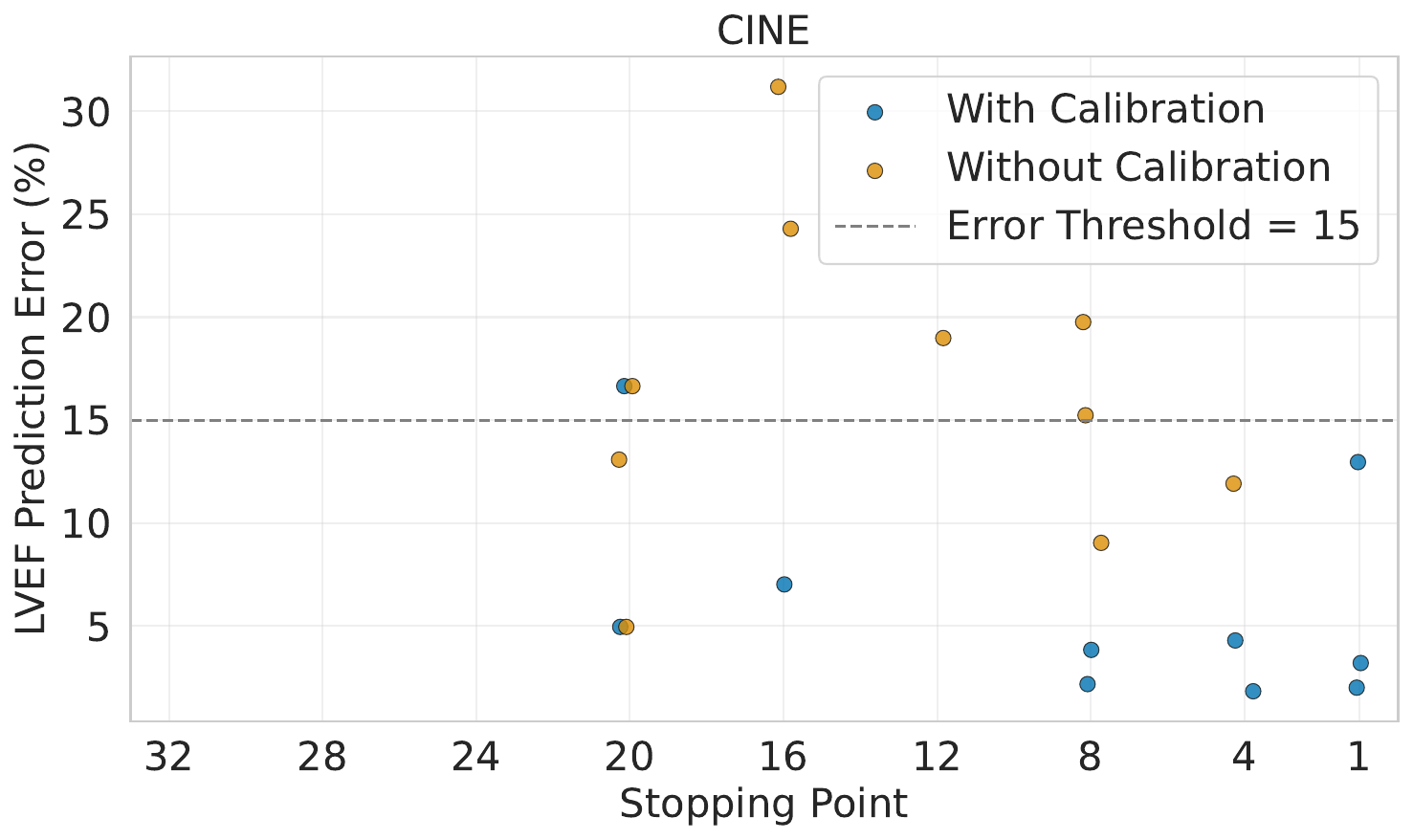}
    \caption{Performance of uncertainty-guided early stopping with and without calibration. The prediction error at point of stopping is plotted against the stopping point determined by our uncertainty criterion. Results are shown for (left) the SKM-TEA dataset, with prediction error measured in $cm^3$, and (right) the CINE dataset, with LVEF prediction error shown in percent. Each point represents a single reconstruction. The model with calibration (blue) reliably terminates acquisition at lower acceleration rates with errors mostly below the task-specific thresholds (dashed lines). In contrast, the uncalibrated model (orange) often produces reconstructions with unacceptable errors while stopping comparably early.}
    \label{fig:error}
\end{figure}

We next analyzed the behavior of the uncertainty-guided stopping mechanism, with quantitative results visualized in Figure~\ref{fig:error}. Our method successfully determines patient-specific scan durations; however, its effectiveness is critically dependent on calibration. Without calibration, the mechanism consistently terminated scans fairly early. For the SKM-TEA dataset, every scan was stopped at the highest acceleration factor (32x), while CINE scans stopped at an average of 13.2x. In contrast, applying conformal calibration resulted in significantly longer scan durations, with average stopping points of 4.35x for SKM-TEA and 8.3x for CINE.

This difference in stopping behavior directly translated to a substantial reduction in prediction error. For the SKM-TEA dataset, the average volume error at stopping decreased from 0.91 $cm^3$ (uncalibrated) to 0.42 $cm^3$ (calibrated). Similarly, for the CINE dataset, the average LVEF error was reduced from 16.5\% (uncalibrated) to 5.90\% (calibrated), underscoring the necessity of calibration for achieving reliable downstream predictions.

We next analyzed the behavior of the uncertainty-guided stopping mechanism. As shown in Figure~\ref{fig:error}, our method successfully determines patient-specific scan durations rather than relying on a fixed acquisition time. To assess the impact of calibration, we compared the distribution of stopping points determined by uncalibrated versus calibrated uncertainties. Without calibration, the mechanism consistently terminated scans fairly early. This was particularly pronounced for the SKM-TEA dataset, where every scan was stopped at the highest acceleration factor (32x). In contrast, applying conformal calibration resulted in significantly longer and more varied scan durations which showed an average stopping at 4.35x for the SKM-TEA dataset. Similarly, for the CINE dataset we observed an average stopping at 13.2x for the uncalibrated and 8.3x for the calibrated case. Additionally, we analyzed the error at stopping which can be seen in Figure \ref{fig:error}. For both datasets, the error at stopping was higher in the uncalibrated compared to the calibrated case. For the SKM-TEA dataset, the average error for the predictions at stopping in the uncalibrated case was 0.91 $cm^3$ and 0.42 $cm^3$ for the calibrated case. For the CINE dataset, the LVEF error for uncalibrated stops was on average 16.5\% and for the calibrated case 5.90\%. 

To evaluate the statistical reliability of the uncertainty intervals at the moment of stopping, we measured the empirical coverage—the percentage of test cases where the ground truth metric fell within the predicted interval. For SKM-TEA, uncalibrated intervals achieved only 17.6\% coverage, which increased to 61.1\% after calibration. For the CINE dataset, coverage improved from 20.0\% to 85.7\% with calibration. While calibration substantially improved reliability, the empirical coverage for both datasets remained below the target of 90\%.

Finally, our method is computationally efficient and suitable for real-time implementation. The entire pipeline—encompassing probabilistic reconstruction (20 samples), segmentation, and calibrated uncertainty estimation—requires approximately 28 ms per slice on an NVIDIA A100 GPU. This translates to an overhead of less than 0.4 seconds for a typical CINE volume and under 4.5 seconds for a full SKM-TEA volume, making the approach practical for online decisions on scan termination.

% \begin{figure}[t]
%     \centering
%     \includegraphics[width=0.49\linewidth]{figs/stopping_distribution_skm.pdf}
%     \includegraphics[width=0.49\linewidth]{figs/stopping_distribution_cine.pdf}
%     \caption{Distribution of how often the scan was terminated for each acceleration factor for SKM-TEA (left) and CINE (right) datasets. The x-axis shows different experimental settings (acceleration factors), and the y-axis indicates the percentage of runs that stopped at each setting. Results are shown separately for models with and without calibration. Notably, models without calibration tend to stop earlier compared to models with calibration for both datasets.}
%     \label{fig:stopping}
% \end{figure}

\subsection{Qualitative Results}
\label{sec:qualitative}
\begin{figure}[t]
    \centering
    \includegraphics[width=0.99\linewidth]{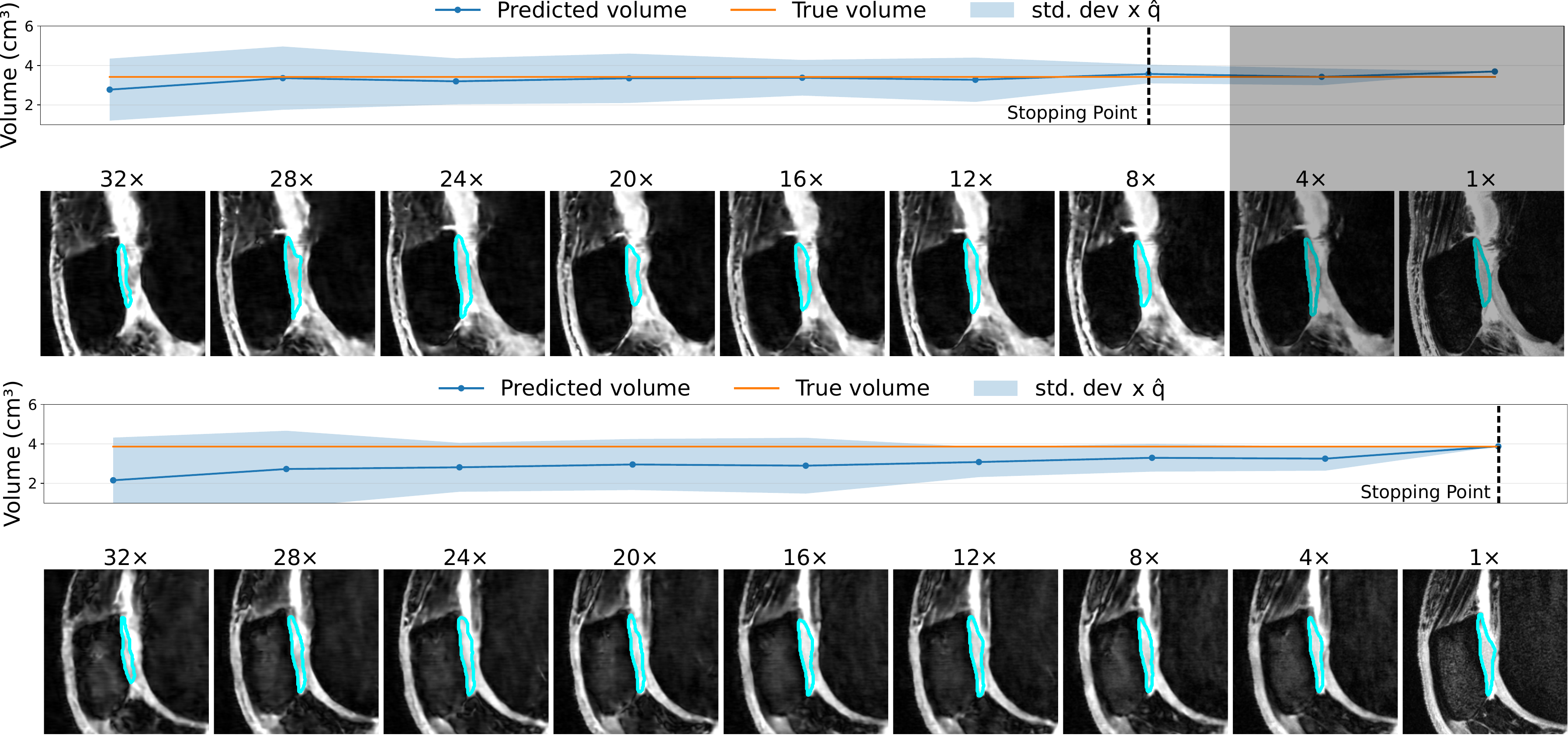}
    \caption{Patellar cartilage volume estimations along with calibrated uncertainty bounds and examples of reconstructions and segmentations for all acceleration factors for the SKM-TEA dataset. The top subject (MTR\_196) displays a case of lower uncertainty (and notably lower error) whereas the bottom subject (MTR\_120) displays higher uncertainty and therefore a longer scan time. The grayed out area indicates the scans that would not have been acquired due to early stopping.}
    \label{fig:skm_unc}
\end{figure}

\begin{figure}[t]
    \centering
    \includegraphics[width=0.99\linewidth]{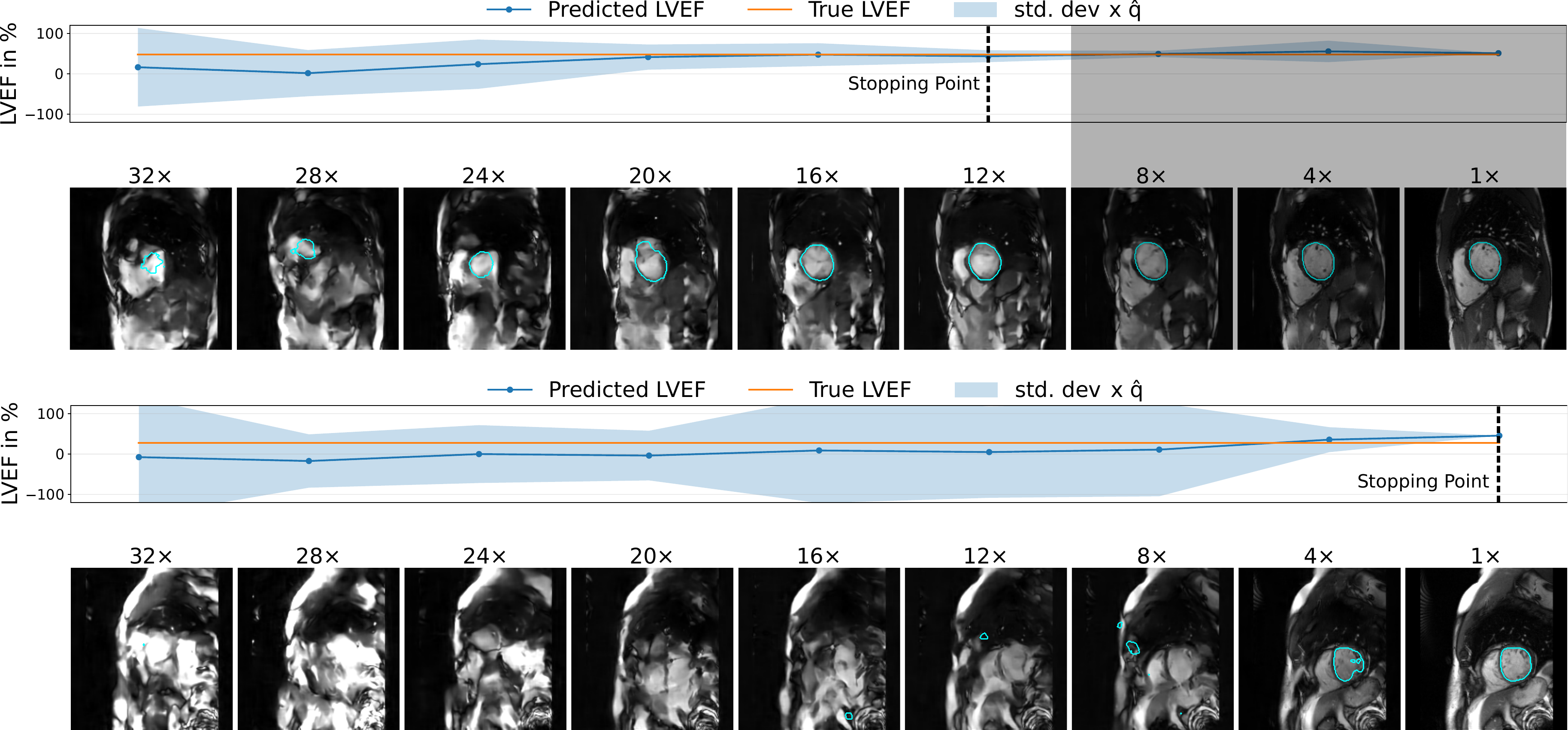}
    \caption{LVEF estimates along with calibrated uncertainty bounds and examples of reconstructions and segmentations for all acceleration factors for the CINE dataset. The top subject displays a case of lower uncertainty whereas the bottom subject displays higher uncertainty. One can clearly see the differences in segmentation quality that lead to the high uncertainty for the lower subject. The grayed out area indicates the scans that would not have been acquired due to early stopping.}
    \label{fig:cine_unc}
\end{figure}

\begin{figure}[t]
    \centering
    \includegraphics[width=0.9\linewidth]{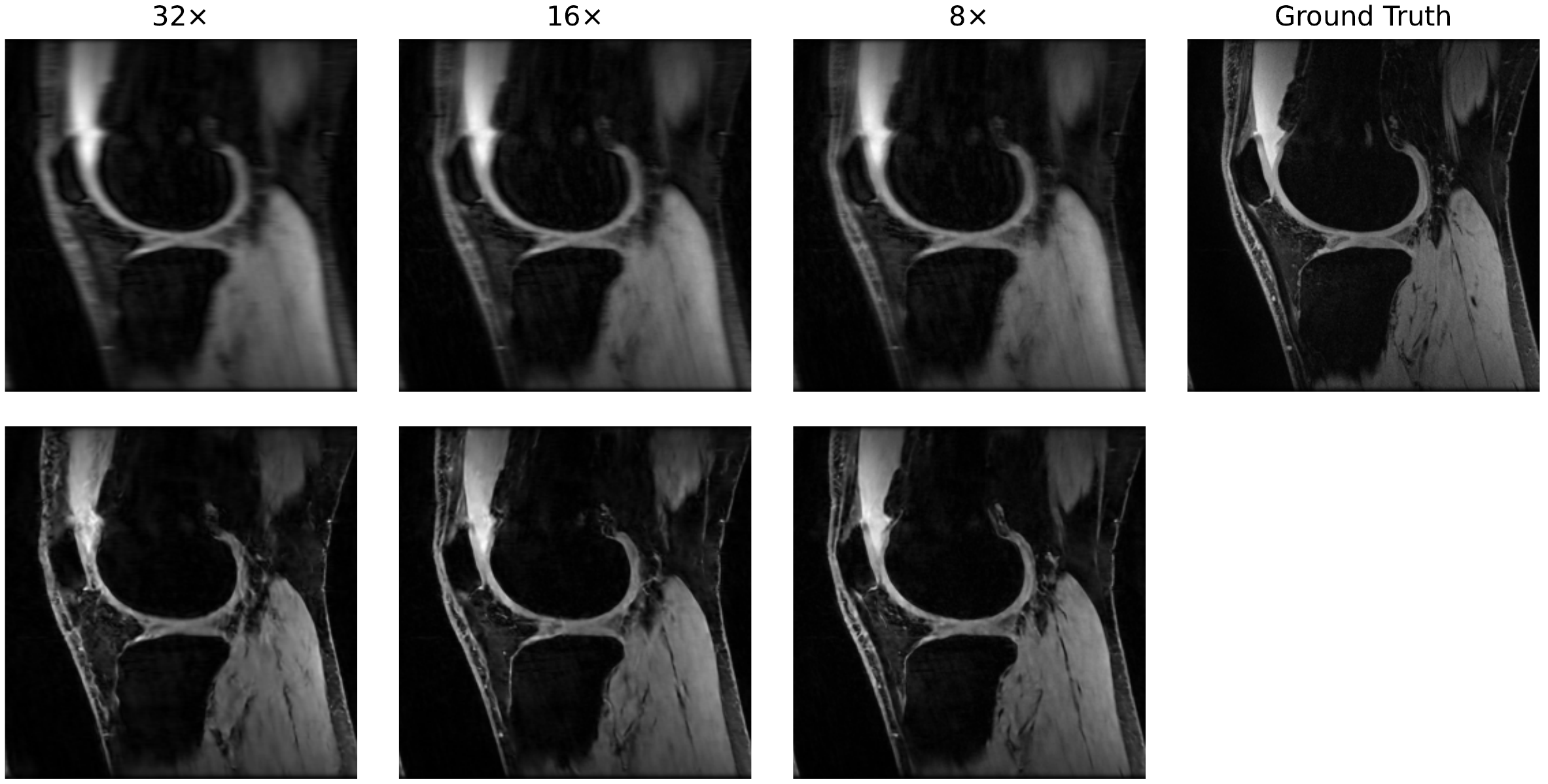} 
    \caption{Example reconstructions for the SKM-TEA dataset. The top row shows the undersampled input images along with the ground truth, and the bottom row shows the corresponding model reconstructions at 32x, 16x, and 8x acceleration.}
    \label{fig:recon_examples_skm}
\end{figure}

\begin{figure}[t]
    \centering
    \includegraphics[width=0.9\linewidth]{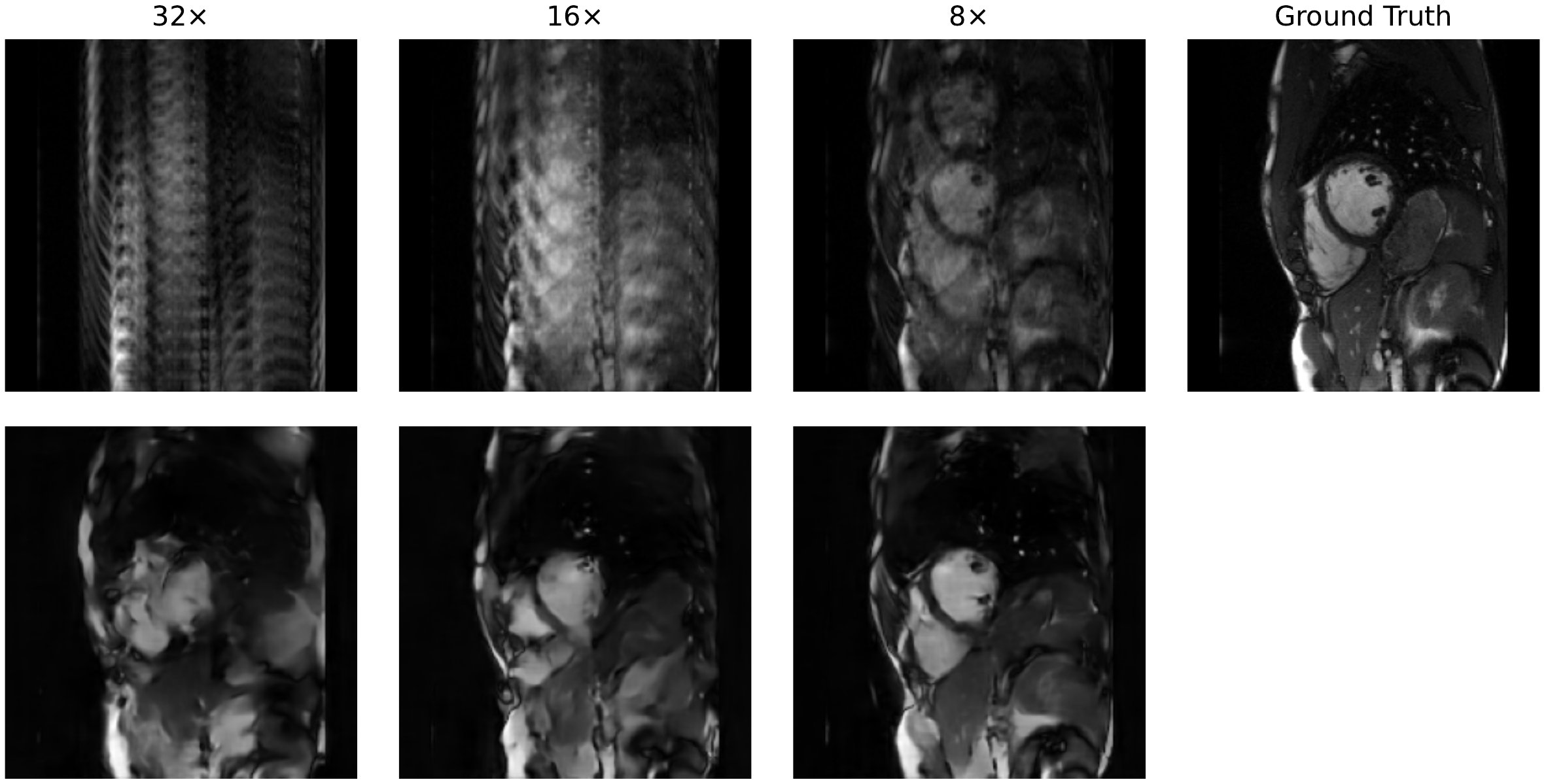}
    \caption{Example reconstructions for the CINE dataset. The top row shows the undersampled input images along with the ground truth, and the bottom row shows the corresponding model reconstructions at 32x, 16x, and 8x acceleration.}
    \label{fig:recon_examples_cine}
\end{figure}

To provide a qualitative understanding of our dynamic stopping mechanism, Figures~\ref{fig:skm_unc} and~\ref{fig:cine_unc} show representative cases of both early and late scan terminations. Each figure visualizes the evolution of the reconstruction, segmentation, and the downstream metric along with its calibrated uncertainty as more k-space data is acquired. As expected, we observe a consistent trend across all examples: as the acquisition progresses, reconstruction quality and segmentation accuracy visibly improve. Moreover, the prediction uncertainty decreases the more k-space data is being collected. Additional reconstruction examples are displayed in Figure~\ref{fig:recon_examples_skm} and~\ref{fig:recon_examples_cine}. Concurrently, the downstream metric estimation converges toward the ground truth value while the corresponding uncertainty bands narrow. Crucially, instances of high uncertainty consistently correspond to visible artifacts, segmentation errors, and larger deviations in the final metric, confirming that our uncertainty estimates effectively track acquisition quality.

\section{Discussion}

Our study demonstrates that downstream uncertainty can effectively guide dynamic MRI scan termination, enabling patient-specific acquisition times. We establish that conformal calibration is indispensable for this task, as uncalibrated uncertainty estimates from deep learning models are systematically overconfident and lead to premature scan termination with unacceptably high error rates. By providing statistically meaningful uncertainty intervals, our calibrated approach offers a robust framework for balancing scan time and diagnostic confidence.

\subsection{Interpretation of Key Findings}
Our results confirm the expected trade-off between acquisition speed and image quality, where both reconstruction and segmentation performance improve with increased k-space sampling. The performance gap between the SKM-TEA and CINE datasets highlights the significant impact of the k-space sampling strategy on task difficulty. To place our results in context, we verified that the performance of our models on fully-sampled data is comparable to benchmarks reported in the original SKM-TEA publication \cite{desai2022skm} and related CINE segmentation work \cite{wang2020ica}, confirming the validity of our underlying models.

The core contribution of this work lies in the dynamic stopping mechanism. The dramatic difference between uncalibrated and calibrated stopping points (Figure~\ref{fig:error}) reveals a critical insight: raw neural network uncertainties are not reliable proxies for model error. The uncalibrated models were consistently overconfident, terminating scans when the downstream metric error was still high (Figure~\ref{fig:error}). This misalignment poses a significant clinical risk. Conformal calibration corrects this by widening the uncertainty intervals to better reflect the true potential for error, leading to more appropriate and safer stopping decisions. This finding aligns with a growing body of literature emphasizing the necessity of calibration for deploying machine learning models in high-stakes medical applications \cite{wundram2024conformal, angelopoulos2022image}.

Furthermore, our qualitative results (Figures~\ref{fig:skm_unc}, \ref{fig:cine_unc}) visually corroborate these quantitative findings. The clear correlation between wider uncertainty bands, visible image artifacts, and inaccurate segmentations provides intuitive evidence that the calibrated uncertainty is a meaningful and trustworthy indicator of quality.

\subsection{Limitations and Future Work}
We acknowledge several limitations in this study. First, our reconstruction model does not enforce data consistency, which could potentially improve image quality and reduce uncertainty, leading to earlier, more efficient scan termination. Integrating a data consistency term within the probabilistic framework is a clear next step.

Second, our framework adapts the scan duration but not the acquisition strategy, as it relies on a discrete set of predefined undersampling masks. A more advanced approach would optimize the k-space trajectory in real-time, selecting the most informative measurements to reduce uncertainty as quickly as possible. This could be achieved using techniques like reinforcement learning or Bayesian experimental design.

Finally, a key challenge lies in the trade-off between the statistical validity and the clinical utility of the uncertainty intervals. While calibration improved reliability, the empirical coverage on our test sets did not consistently achieve the nominal 90\% target, likely due to a distribution shift between the calibration and test data. Statistically, achieving the target coverage would require generating even wider uncertainty intervals. However, intervals that are too wide, while statistically sound, may offer limited clinical value. Setting a stricter stopping criterion to ensure clinical utility would, in turn, result in most scans running to completion, negating the benefit of the adaptive approach. This dilemma reveals that the primary limiting factor is not the calibration method itself, but the predictive performance of the underlying model. Large uncertainty widths are fundamentally a symptom of high prediction error. Therefore, to generate intervals that are both statistically valid and clinically useful, future work must prioritize improving the base predictive accuracy for the metrics of interest, such as LVEF and patellar cartilage volume. This would naturally lead to narrower, more decisive uncertainty bounds.

In summary, the path toward real-world clinical implementation requires addressing these limitations. Future work will focus on integrating data consistency into the reconstruction, developing adaptive k-space sampling strategies, and, most critically, enhancing the core predictive power of our models. By improving model accuracy, we can generate uncertainty estimates that are not only statistically robust but also sufficiently precise to drive meaningful real-time decisions in a clinical scanner.

% Finally, while calibration significantly improved the reliability of our uncertainty intervals, the empirical coverage on the test sets did not consistently meet the 90\% target. This indicates a potential distribution shift between the calibration and test sets. To fulfill the statistical vilidity, these intervals should be even wider. However, the interval width might currently be already too wide in order to have a useful clinical application. For example, an LVEF between 52\% and 72\% is considered normal \cite{sawyer2017encyclopedia} and everything below is considered abnormal. A chosen stopping criterion of 15\% uncertainty band width might be too wide since it would cover the whole range between healthy and pathological. Chosing the stopping criterion tighter would lead to a high amount of full scans and therefore no time is being saved. However, large uncertainty widths indicate that the model predictions are still too bad with high error probability. So in order to make this more clinically applicable, this problem needs to be addressed by improving the prediction of the metrics of interest, in our case LVEF for the CINE dataset. Simlarly for the SKM-TEA dataset and patellar cartilage volume prediction. Future work should address the overall prediction performance as well as providing more clinically acceptable uncertainty interval widths. 

% Ultimately, after addressing the points mentioned above, future work aims for implementing this in actual scanners for real-time decision making.

\section{Conclusion}
Deep learning models have dramatically accelerated magnetic resonance imaging, reducing scan times while preserving diagnostic quality \cite{heckel2024deep}. However, this acceleration is typically based on fixed, pre-determined protocols that are not tailored to the patient or a specific diagnostic question. The central challenge in creating more efficient, patient-specific acquisition protocols is determining the precise moment sufficient k-space data has been acquired for a reliable diagnosis. This requires a real-time signal of data sufficiency, a role that can be filled by quantifying model uncertainty. Despite its potential, leveraging uncertainty to dynamically control the acquisition process and enable early stopping remains a largely unexplored area in clinical imaging pipelines.

Our work addresses this fundamental gap by providing a principled approach for leveraging uncertainty arising during accelerated MR acquisition to determine reliable stopping points. We demonstrate that uncertainty estimates can be effectively used to enable dynamic scan termination, allowing for patient-specific optimization of scan duration. Our methodology is validated across two distinct datasets, and we further enhance the reliability of stopping decisions through uncertainty calibration with mathematical guarantees.

Future work should focus on integrating data consistency into the reconstruction model, enabling adaptive k-space sampling, and improving prediction accuracy to achieve clinically useful uncertainty intervals. These steps are essential for translating the proposed framework into real-world applications and making uncertainty-aware MRI acquisition clinically viable.
% Additional research directions include the incorporation of temporal consistency constraints for dynamic imaging, the development of more sophisticated undersampling strategies that better reflect clinical acquisition patterns, and the exploration of alternative uncertainty quantification methods that may provide more informative estimates for stopping decisions.

\section*{CRediT Authorship Contribution Statement}
\textbf{Paul Fischer:} Writing – review \& editing, Writing – original draft, Visualization, Validation, Software, Methodology, Investigation, Formal analysis, Conceptualization. \textbf{Jan Nikolas Morshuis:} Writing – review \& editing, Conceptualization. \textbf{Thomas Küstner:} Writing – review \& editing, Validation, Data curation, Conceptualization. \textbf{Christian Baumgartner:} Supervision, Writing – review \& editing, Validation, Methodology, Conceptualization, Resources, Project administration. 

\section*{Declaration of Competing Interest}
The authors declare that they have no known competing financial interests or personal relationships that could have appeared to influence the work reported in this paper.

\section*{Acknowledgments}
Funded by the Deutsche Forschungsgemeinschaft (DFG, German Research Foundation) under Germany’s Excellence Strategy - EXC number 2064/1 - Project number 390727645. The authors thank the International Max Planck Research School for Intelligent Systems (IMPRS-IS) for supporting Paul Fischer and Jan Nikolas Morshuis.

% \appendix
% \section{Example Appendix Section}


% %% For citations use: 
% %%       \cite{<label>} ==> [1]

% %%
% Example citation, See \cite{mehta2019propagating}.




\bibliographystyle{elsarticle-num}  % or elsarticle-harv, elsarticle-num-names
\bibliography{bibliography}
\end{document}